\documentclass[12pt]{article}

\usepackage{amsmath,amssymb}
\usepackage{authblk}
\title{On nonadditive anisotropic relativistic hydrodynamics}
\date{}
\author{A.V. Leonidov}

\affil{{\small P.N. Lebedev Physical Institute, Moscow, Russia }}

\begin{document}
\maketitle

\begin{abstract}
Non-additive generalisation of relativistic anisotropic anisotropic hydrodynamics is described. In the particular case of 0+1 boost-invariant hydrodynamics additional entropy production due to non-additivity is calculated. 

\end{abstract}

Developing a consistent physical picture of ultrarelativistic heavy ion collisions still remains an outstanding challenge, see e.g.  a recent concise summary of some of its relevant  aspects in \cite{Gelis:2021zmx}. The focus of the present letter is on constructing a theoretical description of anisotropic strongly interacting matter created in such collisions. 

A fundamental origin of strong momentum anisotropy of matter created at early and intermediate stages of high energy heavy ion collisions is believed to be in the strong momentum anisotropy of glasma \cite{Lappi:2006fp}, the dense gluon system created at its early stages. A direct way of accounting for this momentum anisotropy is in turning to  anisotropic hydrodynamics, see e.g. a recent review \cite{Alqahtani:2017mhy}. Below we will closely follow an approach of constructing such anisotropic hydrodynamics from kinetic theory discussed in \cite{Martinez:2010sc,Martinez:2009ry}\footnote{For a detailed discussion see \cite{Molnar:2016vvu}.}. Momentum anisotropy results in new interesting effects such as, e.g., modification of the Mach cone \cite{Kirakosyan:2018afm}.

A more subtle feature is the fact that the matter under discussion is dense and strongly interacting. On fundamental grounds one expects, in particular, that that energy and entropy characterising such strongly interacting systems can not be additive thus making it necessary to generalise the usual statistical physics approach to their description. A natural possible generalisation is a non-additive formulation of statistical mechanics \cite{Tsallis:2009}. Implications of this formalism to particle production in heavy ion collision, in particular to transverse momentum spectra, is discussed, in particular, in \cite{Cleymans:2013rfq,Cleymans:2017kzp,Shen:2019kil,Biro:2020kve}. A non-additive generalisation of hydrodynamics based on the corresponding generalisation of the Boltzmann kinetic equation \cite{Lavagno:2002hv} was constructed in \cite{Biro:2011bq,Biro:2012ix}\footnote{An alternative construction was developed in \cite{Osada:2008sw,Osada:2008da}.}.

As our aim is in developing an approach taking into account both momentum anisotropy and non-additivity, to construct non-additive anisotropic hydrodynamics we need to consider a non-additive  Boltzmann kinetic equation for a distribution function characterised by momentum anisotropy.  In the relaxation time approximation considered below it reads 
\begin{equation}\label{beq}
p^\mu \partial_\mu \left[ f(x,p)^q \right] = - \frac{p^\mu u_\mu}{\tau_{\rm eq}}  \left[ f^q (x,p,\xi \vert \Lambda) - f^q_{\rm eq} (x,p \vert \Lambda_{\rm eq}) \right]
\end{equation}
where $q$ is a parameter of the Tsallis distribution controlling the degree of non-additivity (see below equation\eqref{fts}), $\tau_{\rm eq}$ is a relaxation scale and the anisotropic distribution $f (x,p, \xi \vert \Lambda)$ is assumed to have the  Romatschke-Strikland \cite{Romatschke:2003ms} form 
\begin{equation}\label{fRS}
 f (x,p,\xi \vert \Lambda)  =  f_{\rm iso} \left( \frac{{\bf p}^2+\xi p^2_z}{\Lambda^2} \right) 
\end{equation}
where, in turn, $\xi(t)$ and $\Lambda(t)$ are time-dependent parameters determining the  degree of momentum anisotropy and the magnitude of the momentum scale correspondingly and we have assumed that the distribution function is anisotropic only in longitudinal direction. The function
\begin{equation}\label{feq}
 f_{\rm eq}(x,p, \vert \Lambda_{\rm eq})  =  f_{\rm iso} \left( {\bf p}^2 / \Lambda_{\rm eq}^2 \right)
\end{equation}
corresponds to an equilibrated isotropic state characterised by an effective scale $\Lambda_{\rm eq}$. For the non-additive kinetic formalism under consideration the equilibrium distribution 
$f_{\rm eq} (x,p \vert \Lambda_{\rm eq})$ is of a Tsallis form  
\begin{equation}\label{fts}
 f_{\rm eq} (x,p) = \left[1-(1-q) {\bf p}^2/\Lambda_{\rm eq}^2  \right]^{1/(1-q)},
\end{equation} where in the limit $q \to 1$ one recovers the usual Boltzmann distribution.

The key quantities used in constructing non-additive hydrodynamics from kinetic formalism are generalised particle current $N^\mu$, energy-momentum tensor $T^{\mu \nu}$ and entropy current $S^\mu$:
\begin{eqnarray}
N^\mu & = &  \int \frac{d^3 p}{(2 \pi)^3 p^0} p^\mu f(x,p)^q \label{Nmunu} \\
T^{\mu \nu} & = & \int \frac{d^3 p}{(2 \pi)^3 p^0} p^\mu p^\nu f(x,p)^q \label{Tmunu} \\
S^\mu & = & - \int \frac{d^3 p}{(2 \pi)^3 p^0} p^\mu \left[ f(x,p)^q  \ln_q   f(x,p) -  f(x,p) \right] \label {Smu}
\end{eqnarray}
where
\begin{equation}
\ln_q (x) = \frac{x^{1-q}-1}{1-q}
\end{equation}
Let us note that in the literature  \cite{Lavagno:2002hv,Biro:2011bq,Biro:2012ix} one can find different suggestions for the generalised entropy current. The expression in  \eqref{Smu} follows the choice made in \cite{Biro:2011bq,Biro:2012ix}.

To ensure the energy-momentum conservation in \eqref{beq} we employ the Landau matching condition, see e.g.  \cite{Alqahtani:2017mhy,Biro:2011bq,Biro:2012ix} for the energy $\epsilon=T^{00}$
\begin{equation}\label{Lmc}
\epsilon(\xi,\Lambda) = \epsilon_{\rm eq}(\Lambda_{\rm eq}),
\end{equation}
where calculations in the left- and right-hand side of \eqref{Lmc} are performed with the distribution functions \eqref{fRS} and \eqref{feq} correspondingly leading to the following matching condition on $\Lambda_{\rm eq}$ and $\Lambda$:
\begin{equation}
\Lambda_{\rm eq} = {\cal R(\xi)}^{1/4} \Lambda, \;\;\;  {\cal R}(\xi)  = \frac{1}{2} \left( \frac{1}{1+\xi} + \frac{\arctan \sqrt{\xi}}{\sqrt{\xi}}\right)
\end{equation}

In this letter, following \cite{Biro:2011bq,Biro:2012ix}, we consider the boost-invariant  $0+1$ - dimensional hydrodynamics in which all quantities depend only on the proper time $\tau$ defined by Milne coordinates $(\tau,\eta)$ defined by 
\begin{equation}
t  =  \tau \cosh \eta, \;\;\; z = \tau \sinh \eta
\end{equation}

Hydrodynamic equations are those for the first two moments of the Boltzmann equation for particle and energy-momentum currents (\ref{Nmunu},\ref{Tmunu}). It is easy to see that the correspoding calculations for the non-additive case closely follow those described in  \cite{Martinez:2010sc} and result in the same evolution equations for $\Lambda$ and $\xi$:
\begin{eqnarray}\label{evxilam}
 \partial_\tau \xi & = & \frac{2 (1+\xi) }{\tau} - \frac{4  (1+\xi)}{\tau_{\rm eq}} {\cal R}(\xi) {\cal G} (\xi) \nonumber \\
 \partial_\tau  \Lambda & = & \frac{1+\xi}{\tau_{\rm eq}} {\cal R}'(\xi) {\cal G} (\xi) \Lambda
\end{eqnarray}
where
\begin{equation}
{\cal G} (\xi) = \frac{{\cal R}^{3/4}(\xi) \sqrt{1+\xi}-1}{2{\cal R}(\xi) +3(1+\xi) {\cal R}'(\xi)}
\end{equation}
Let us stress, that the number density \eqref{Nmunu} and energy-momentum tensor \eqref{Tmunu} in the non-additive case do of course differ from their additive counterparts.

Let us turn to the analysis of the evolution of the entropy density $S \equiv S^0$. In performing the calculation  it is convenient to explicitly separate the dependence on the anisotropy parameter
\begin{equation}
S (\xi,\Lambda) = \frac{1}{\sqrt{1+\xi}} S_{\rm iso} (\Lambda)
\end{equation}
where
\begin{equation}
S_{\rm iso} (\Lambda)= \frac{ \Lambda^3}{(2 \pi)^2} \int d w \; w^{1/2} \left[f^q_{\rm iso}(w) \ln_q f_{\rm iso}(w) - f_{\rm iso}(w) \right]
\end{equation}
and $w = p^2/\Lambda^2$. Let us write the equation for $\partial_\tau S$ in the following form:
\begin{equation}
\partial_\tau S = \Delta_0 (\tau) + (q-1) \Delta_q (\tau)
\end{equation}
where we have separated the additive $\Delta_0 (\tau)$ and non-additive  $(q-1) \Delta_1 (\tau)$ such that in the additive limit $q \to 0$ we are left with the first term only.
We have
\begin{eqnarray}
 \Delta_0 (\tau) & = & - \left[ \frac{1}{2} \frac{1}{1+\xi} ( \partial_\tau \xi )  -  \frac{3}{\Lambda} ( \partial_\tau \Lambda ) \right] S \nonumber \\
 \Delta_q (\tau) & = &  \frac{1}{\sqrt{1+\xi}} \frac{2}{\Lambda} \partial_\tau \Lambda \int dw \; w^{1/2} \; {\rm Ln}_q f_{\rm iso} (w)
\end{eqnarray}
where
\begin{equation}
{\rm Ln}_q \; f_{\rm iso} (w) \equiv q \; \frac{f^q_{\rm iso} (w) -1}{q-1}
\end{equation}
Using equations \eqref{evxilam} we finally obtain
\begin{eqnarray}
\Delta_0 (\tau) & =  & \frac{1}{\tau_{\rm eq}} \left[ {\cal R}^{3/4} (\xi) \sqrt{1+\xi} -1\right] S \label{del0} \\
\Delta_q (\tau) & =  & \sqrt{1+\xi} \frac{1}{\tau_{\rm eq}} {\cal R}'(\xi) {\cal G} (\xi)  \int dw \; w^{1/2} \; {\rm Ln}_q f_{\rm iso} (w) \label{del1}
\end{eqnarray}
Equations (\ref{del0},\ref{del1}) present the main result of the paper: non-additivity in describing collective properties of the anisotropic hydrodynamics results in additional  contribution to entropy production described by \eqref{del1} on top of the previously known \cite{Martinez:2010sc} contribution from momentum anisotropy described by the equation \eqref{del0}.

The work was supported by the RFBR project 18-02-40131.

\end{document}